\documentclass[twocolumn]{aastex63}
\shorttitle{Lithium in Open Clusters}
\shortauthors{Twarog}

\begin{document}

\def\teff{$T\rm_{eff }$}
\def\k{$\mathrm {km s}^{-1}$}

\title{Intermediate-to-Low Mass Stars in Open Clusters and the Evolution of Li}

\author{Bruce A. Twarog}
\affiliation{Department of Physics and Astronomy, University of Kansas, Lawrence, KS 66045-7582, USA}
\email{btwarog@ku.edu}

\author{Barbara J. Anthony-Twarog}
\affiliation{Department of Physics and Astronomy, University of Kansas, Lawrence, KS 66045-7582, USA}
\email{bjat@ku.edu}

\author{Constantine P. Deliyannis}
\affiliation{Department of Astronomy, Indiana University, Bloomington, IN 47405-7105, USA}
\email{cdeliyan@indiana.edu}

\author{Aaron Steinhauer}
\affiliation{Department of Physics and Astronomy -- SUNY Geneseo,
Geneseo, NY 14454, USA}
\email{steinhau@geneseo.edu}

\begin{abstract}
Open clusters (OC) of 1-3 Gyr age contain intermediate-to-low-mass stars 
in evolutionary phases of multiple relevance to understanding Li evolution. 
Stars leaving the main sequence (MS) from the hot side of the Lithium dip (LD) at a fixed age can 
include a range of mass, varying degrees of core degeneracy, and  
helium ignition under quiescent or flash conditions. An ongoing survey of a 
significant sample of stars from the giant branch to below the LD in key 
open clusters has revealed patterns that supply critical clues 
to the underlying source of Li variation among stars of differing mass and age.  
While the LD is well established in OC of this age, stars on the 
hot side of the LD can exhibit Li ranging from the apparent primordial 
cluster value to upper limits similar to those found at the LD center, 
despite occupying the same region of the color-magnitude diagram (CMD). Stars 
on the first-ascent giant branch show a dramatic decline in measurable Li that 
correlates strongly with increasing age and reduced turnoff mass. We 
discuss how these trends can be explained in the context of the existence 
of the LD itself and the temporal evolution of individual stars.
\end{abstract}

\section{Introduction}
A perennial justification for studying OC is minimization of the 
inhomogeneity of internal and external parameters within the OC, 
ideally reduced to variation in one fundamental property: mass. All things 
being equal, comparison of trends with mass for OC of differing age 
supplies insight into the means by which internal evolution of a star 
can temporally alter its surface appearance, making the unseen 
visible. OC Li analyses are no exception to this trend and, in fact, 
may be the best direct example of the use of a surface abundance to probe 
atmospheric alteration of stars of varying mass, both with time and 
evolutionary state.

The role of Li among stars of cooler \teff\ /lower mass has been discussed 
by others at this conference and will not be 
repeated.\footnote{This paper was presented at the conference, ``Lithium in the Universe: to Be or not to Be", sponsored by INAF-OAR, Rome, Italy, Nov. 18-22, 2019.}
 We will focus on stars in the 1-2.5 M$_{\sun}$ mass range 
for which the optimal observation window occurs within OC older than 
the Hyades but significantly younger than halo globulars. An exact age 
range remains indeterminate because transition points between the phenomena 
controlling Li evolution on and off the MS are dominated 
first by \teff\  and second by mass \citep{ch01, cu12}. Since the main 
sequence mass of a star of given \teff\  changes with metallicity, stars 
of a given mass but lower metallicity will be altered internally in a 
different manner and at a different age from their metal-rich counterparts.

While not one of our two core topics, it should be noted that OC in the 
age range of interest have drawn attention from Li aficionados due to 
an intriguing pattern tied as much to cosmology as stellar structure. 
Moving up the MS from lower to higher mass/\teff, the time 
rate of Li depletion declines, producing a steepening gradient of Li 
with \teff\ with age in older OC. However, stars in the MS 
\teff\  range between 6500K and 6900K, where Li depletion 
should be nonexistent, populate the LD where Li undergoes a 
dramatic decline relative to stars outside the \teff\  boundary 
\citep{bo86}. This anomaly creates a Li plateau among stars cooler 
than 6000K but hotter than 5500K, a plateau that declines in A(Li) 
with increasing age, illustrated schematically in Fig. 1. Crucially, 
for older OC independent of metallicity, the A(Li) value approached 
by the plateau is 2.4-2.6, tantalizingly close to the Spite plateau 
among halo dwarfs (see, e.g. \citet{ra09, cu12}). The initial A(Li) 
among OC is believed to be positively correlated with [Fe/H]. But, 
OC ranging in [Fe/H] from +0.4 to -0.4 generate plateau dwarfs 
with virtually identical asymptotic A(Li) limits, indicating that all 
dwarfs in this \teff\  range, including halo dwarfs, gradually deplete 
Li to a similar A(Li) value for reasons that remain, at best, unclear. 
The OC failure to reach a value closer to A(Li) $\sim$ 2.2 may simply 
be due to the invariably younger age of the oldest OC (5-8 Gyr) relative 
to the halo (11-13 Gyr). 

\section{Two Key Questions}
Returning to the two current topics of interest, we refer again to Fig. 1 ({\it l}),
with a schematic CMD for an intermediate-age 
OC ($\sim$1.5 Gyr; [Fe/H] = 0.0). With the exception of the subgiant 
and giant branch (green), the corresponding representative trends for A(Li) with
mass are identified by color ({\it r}).

\begin{figure}
\figurenum{1}
\includegraphics[width=\columnwidth]{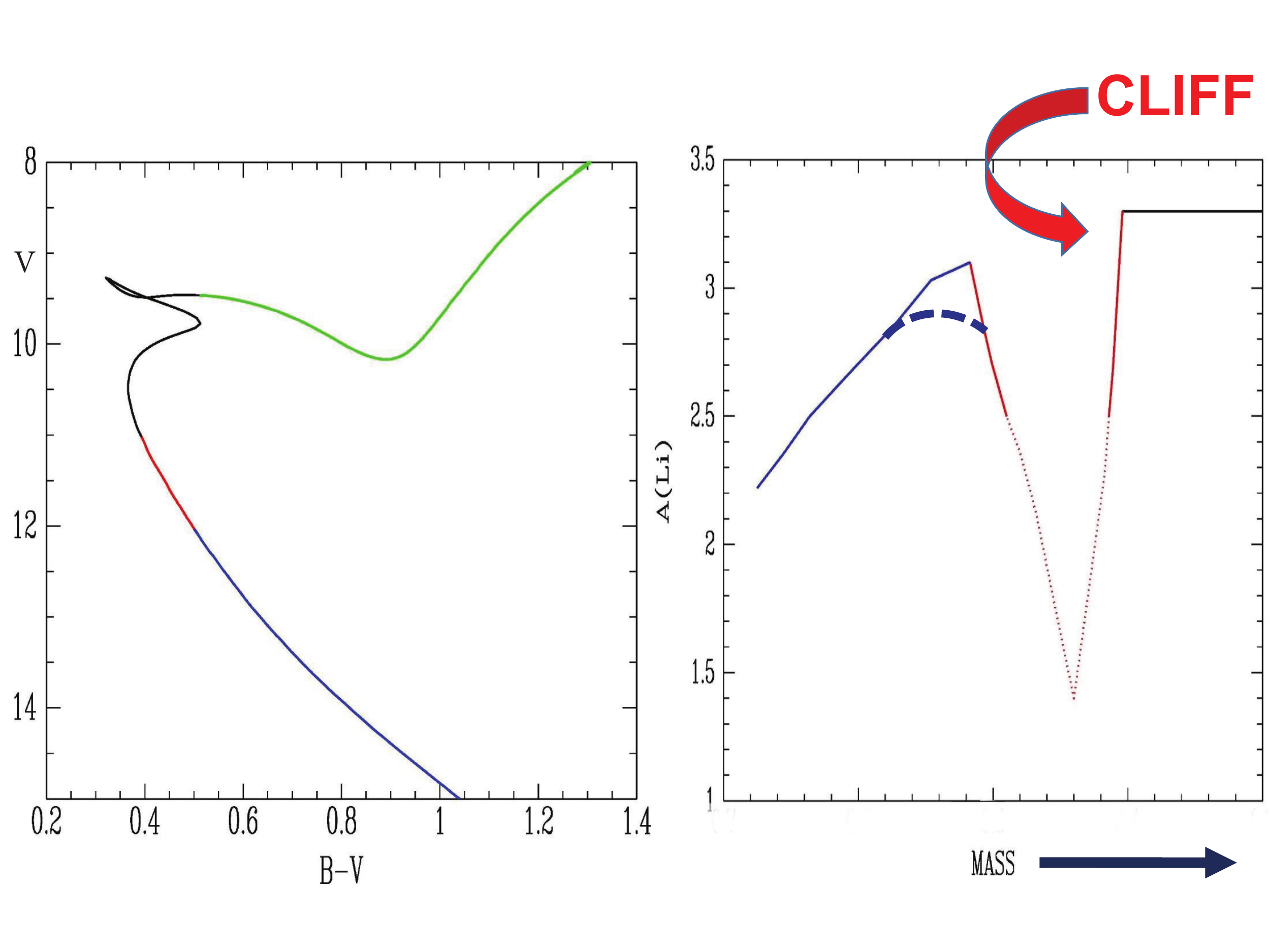}
\caption{({\it l}) Schematic CMD for 1.5 Gyr OC; ({\it r}) Li trend with mass at the turnoff; dashed
curve shows change in the plateau with increased age.}
\end{figure}

The Li plateau level will decline with age ({\it r}-dash). 
Within the LD (red), depth may grow deeper with time. 
However, A(Li) is already so low by 0.65 Gyr that detection of 
further decline with age is technically unfeasible in currently 
studied OC. What is striking is the sharp decline over a very narrow 
mass change at the high-mass LD boundary, designated 
as the {\it cliff} in Fig. 1. Stars above the cliff (black) supposedly 
undergo no Li depletion and should retain the primordial cluster Li. One 
rationale for the study of these cliff dwellers has been to 
highlight the decline in A(Li) experienced by stars on the cool Li 
plateau through comparison with these stars as exemplars of the 
original cluster A(Li). However, in defining the parameters that 
could affect the LD, the sharp edge of the cliff offers an ideal 
marker for tagging the LD boundary, moreso than the continually 
evolving, sloped cool plateau and/or the ill-defined center of 
the LD; both have been used to evaluate the evolution of 
the LD \citep{at09, cu12}. This marker clearly disappears once an 
OC approaches an age where stars leaving the MS for 
the giant branch come from within the LD, as exemplified 
by NGC 6253 \citep{at10, cu12} and M67 \citep{ca16}. For our 
first question, we try to determine if there is evidence that 
the cliff boundary is affected by any property 
other than cluster metallicity, i.e. does the cliff edge shift to higher 
stellar mass as an OC ages?

The second question focuses on the evolution of cliff dwellers 
across the subgiant branch and up the giant branch. Since all 
models predict that post-main-sequence evolution should lead to a 
decline in A(Li) through deepening convection zones and other 
structurally-induced forms of mixing, the distribution of A(Li) 
among evolved stars is a byproduct of these processes acting 
on the initial Li abundance as defined by the stellar value as it 
leaves the MS. Do cliff dwellers retain the actual 
cluster primordial value or are there processes that can alter the 
base value prior to leaving the MS, supplying some insight 
into the source of anomalies like the LD itself?

To answer the first question, we assemble an OC sample of 
comparable metallicity ([Fe/H] = -0.03 to -0.08) but differing age. 
(Note that all OC discussed have been processed for membership 
using radial velocities and Gaia astrometry, and for age 
and distance using isochrones modified by well-defined 
reddening and metallicity.) The OC of interest were, 
until recently, NGC752 \citep{tw15} and NGC3680 \citep{at09}
with ages of 1.45 and 1.75 Gyr, respectively. These OC have 
numerically less-than-rich populations and incomplete Li surveys.  
Merger of the data allows determination of the approximate LD 
location, but exact boundaries remain fluid. Among cooler evolved stars, 
no subgiants survive and the giant sample is so poor that 
classification as first-ascent giants versus clump stars becomes 
an exercise in personal bias, often defined by the presence/absence 
of measurable Li.

To overcome this, a program of high-dispersion spectroscopic 
analyses of extensive samples of stars in rich OC was 
initiated, with NGC6819 (age=2.25 Gyr) among the first where the 
sample from the tip of the giant branch to below the LD 
numbered over 330 stars \citep{de19}. Fig. 2 shows the A(Li) trend 
with mass among stars (blue) at the OC turnoff not classified as 

\begin{figure}
\figurenum{2}
\includegraphics[width=\columnwidth]{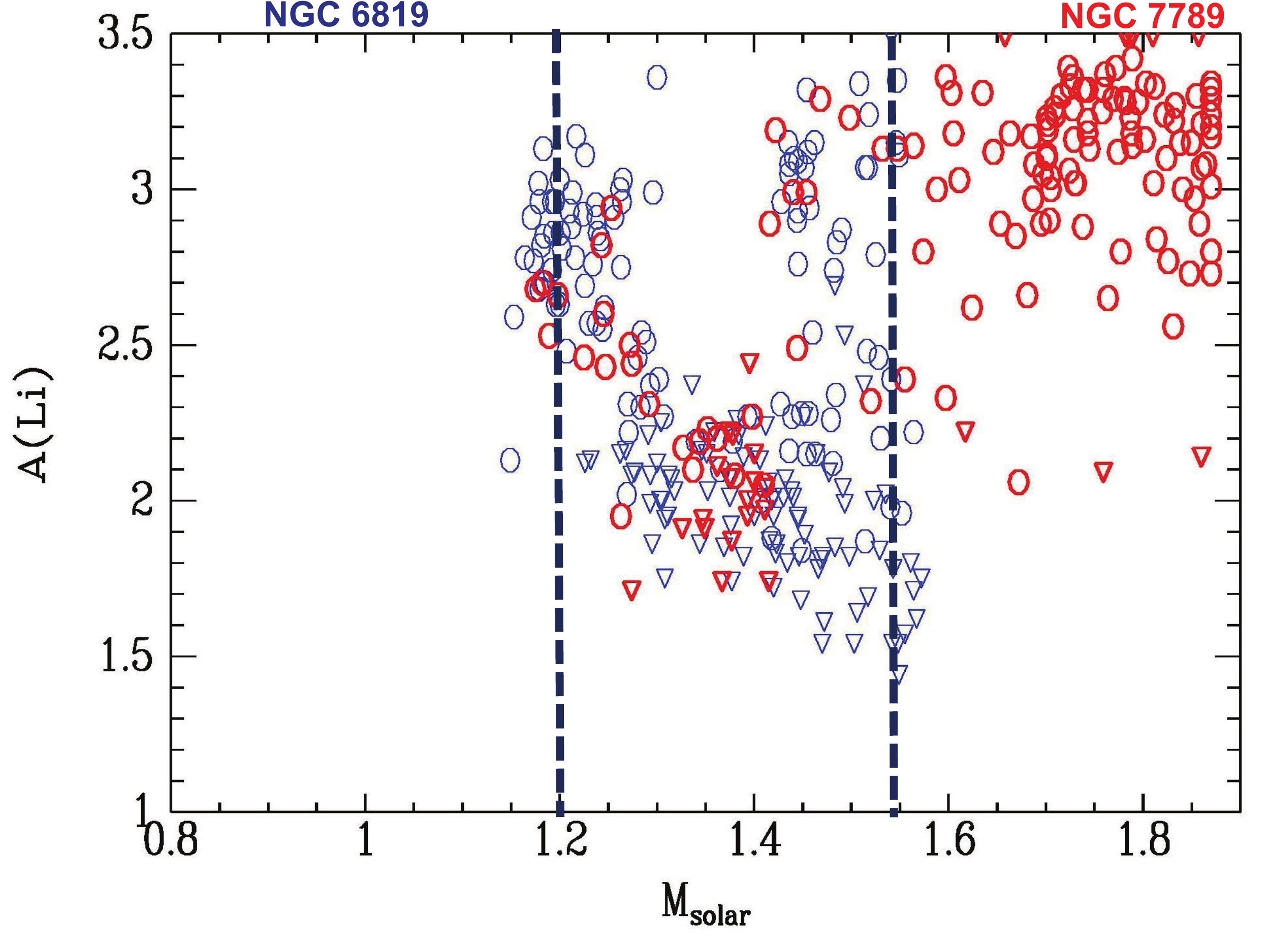}
\caption{LD within NGC6819 (blue) and NGC7789 (red). (Triangles = upper limits). 
Dashed lines delineate the LD from composite sample of NGC752/NGC3680.}
\end{figure}

subgiants; triangles illustrate stars with only A(Li) upper limits. 
Note how sharp the hot boundary is at 1.43 M$_{\sun}$. The vertical 
dashed lines show an approximate estimate of the LD boundaries 
derived using the sample from NGC752/NGC3680. Is the boundary 
difference real or a byproduct of an inadequate sample from the 
composite OC? Fig. 2 shows the A(Li) -- mass trend for the rich OC, NGC7789 (red)
(age=1.5 Gyr), including over 300 stars from the giant branch to 
below the LD. The greater mass range at the turnoff 
identifies NGC7789 as the younger OC but the cliff boundary is 
identical within the uncertainties to that for NGC6819, 
implying that the potentially broader LD for the NGC752/NGC3680 
data is due to the statistical inadequacy of the sample.  

Before addressing the second primary question, an 
issue tied to cliff mass variation with age should be noted. 
Is there significant variation of the cliff 
with metallicity, as expected if the boundary is defined 
on the unevolved MS solely by \teff ? A simple 
comparison between the Hyades/Praesepe boundary at 1.495 M$_{\sun}$ 
for [Fe/H]=+0.15 and the above 1.425 M$_{\sun}$ for [Fe/H]=-0.05 
implies a slope of 0.35 M$_{\sun}$/dex. However, our analysis of 
the metal-poor OC ([Fe/H]=-0.5, age=3.5 Gyr), NGC2243, shows that the turnoff 
just reaches the extant cliff with a boundary at 1.225 M$_{\sun}$. 
A linear fit implies a slope of 0.4 M$_{\sun}$/dex, as found in 
earlier work \citep{at09, cu12}.

We now look at the stars populating the cliff since 
these stars become the foundational source for those occupying 
the subgiant/giant branch. We start with our youngest/richest OC, 
NGC7789. Because it is young, stars leaving the turnoff do so without a 
significant degree of core degeneracy, leading to a poorly 
populated subgiant branch/Hertzsprung gap due the rapid phase 
of evolution after core H-exhaustion. Only a handful 
of first-ascent giants lie below the level of the clump. 

To better understand this critical phase, a Li study 
of NGC2506 ([Fe/H]=-0.3,age=1.8 Gyr) was done. 
Because of lower [Fe/H], the cliff boundary occurs at a 
lower mass than in NGC7789, implying that it will more 
closely resemble the NGC7789 Li turnoff when NGC2506 stars 
of lower mass occupy this phase, i.e. when NGC2506 is older. 
Because the cluster turnoff is populated by lower mass stars, 
the degree of core degeneracy is greater at H-exhaustion and the 
distribution of first-ascent giants is more heavily weighted toward 
the subgiant branch and vertical red giant branch below the 
clump, as confirmed by the its CMD \citep{at18}. Fig. 3 presents 

\begin{figure}
\figurenum{3}
\includegraphics[width=\columnwidth]{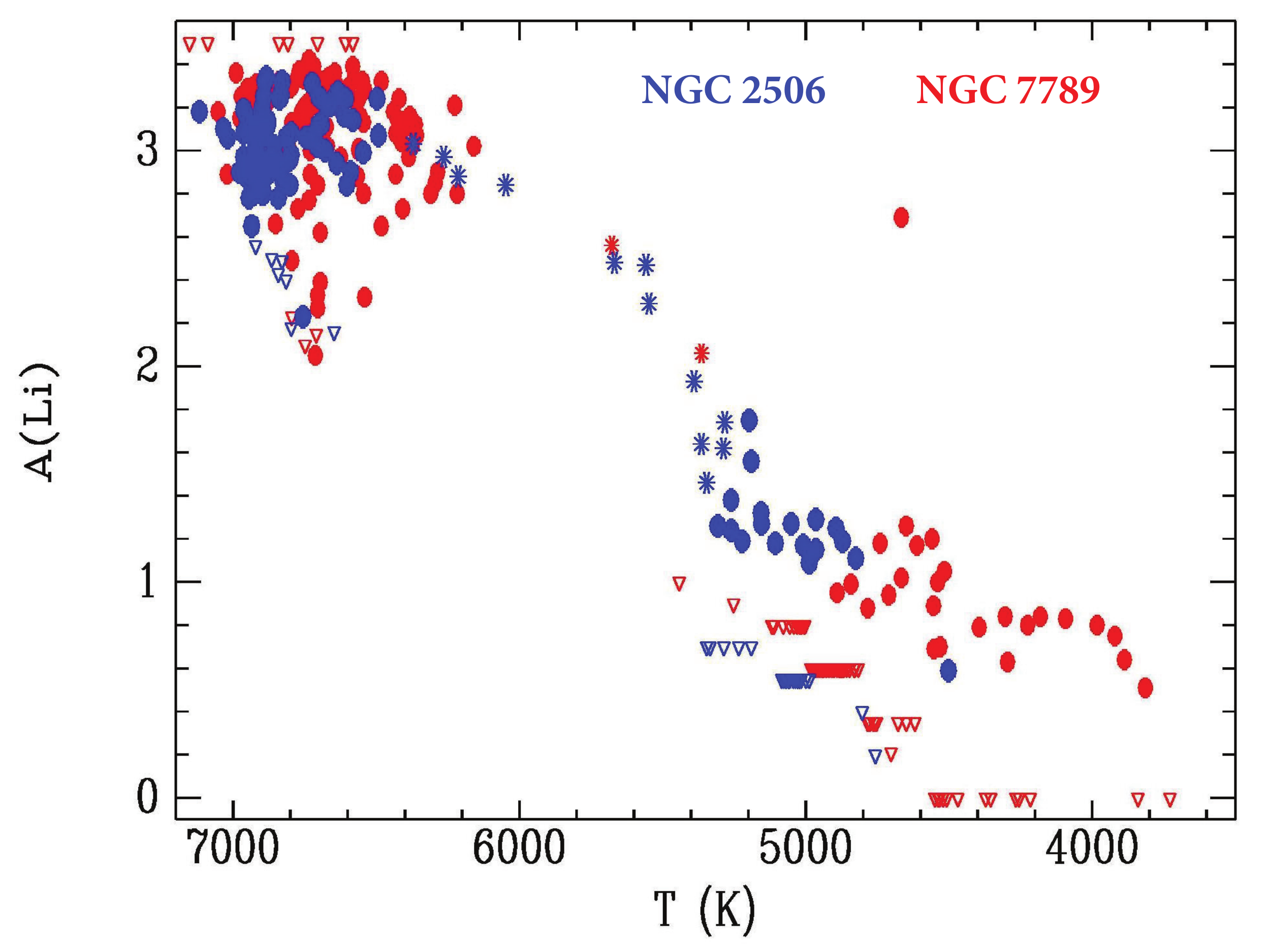}
\caption{Li evolution (NGC2506-blue; NGC7789-red) from above the cliff through the
giant branch (triangles-upper limits; asterisks-subgiants). }
\end{figure}

the Li-\teff\ relation for the two OC from above 
the LD through the giant branch tip
for stars in NGC7789 (red) and NGC2506 (blue). Triangles are 
upper limits, but asterisks are stars classified as {\it subgiants}; 
they map evolution across the subgiant to the base 
of the first-ascent giant branch.

A number of points should be noted: a) by the time the 
stars reach the base of the RGB, A(Li) has been reduced to 
1.5 or less, i.e. below the boundary for Li-rich classification. 
Only one red giant in NGC7789 fits that category \citep{pi84}; 
b) there is evidence for a weak decline 
in A(Li) for stars ascending the giant branch until \teff\ $\sim$4500K, 
potentially the site of the RGB bump, 
after which A(Li) approaches a limiting value of 0.6; c) with only 
a pair of exceptions, no clump stars exhibit measurable 
Li, with a limiting value at 0.8 or less. The 
implication is that He-ignition at the RGB tip 
leads to significant Li depletion, whether done in quiescent 
(NGC7789) or flash (NGC2506) form; d) while there are some cliff 
dwellers with significant Li depletions prior to entering 
the subgiant branch, the majority of turnoff stars at this stage 
lie within the A(Li)= 3.3 to 2.7 range, close to their 
likely initial primordial cluster value.

What happens if we push the OC to a greater age, allowing 
more time for the evolution of the stars on the cliff? For this 
purpose, we combine the data for NGC6819 and NGC2243. 
While NGC2243 is significantly older, it is also significantly 
more metal-poor, producing an evolutionary distribution of stars, 
from the Li standpoint, that should be similar to that in NGC6819, 
i.e. cliff dwellers occupy a very narrow mass band at the top 
of the turnoff.

The NGC6819/2243 equivalent of Fig. 3 is shown in Fig. 4.

\begin{figure}
\figurenum{4}
\includegraphics[width=\columnwidth]{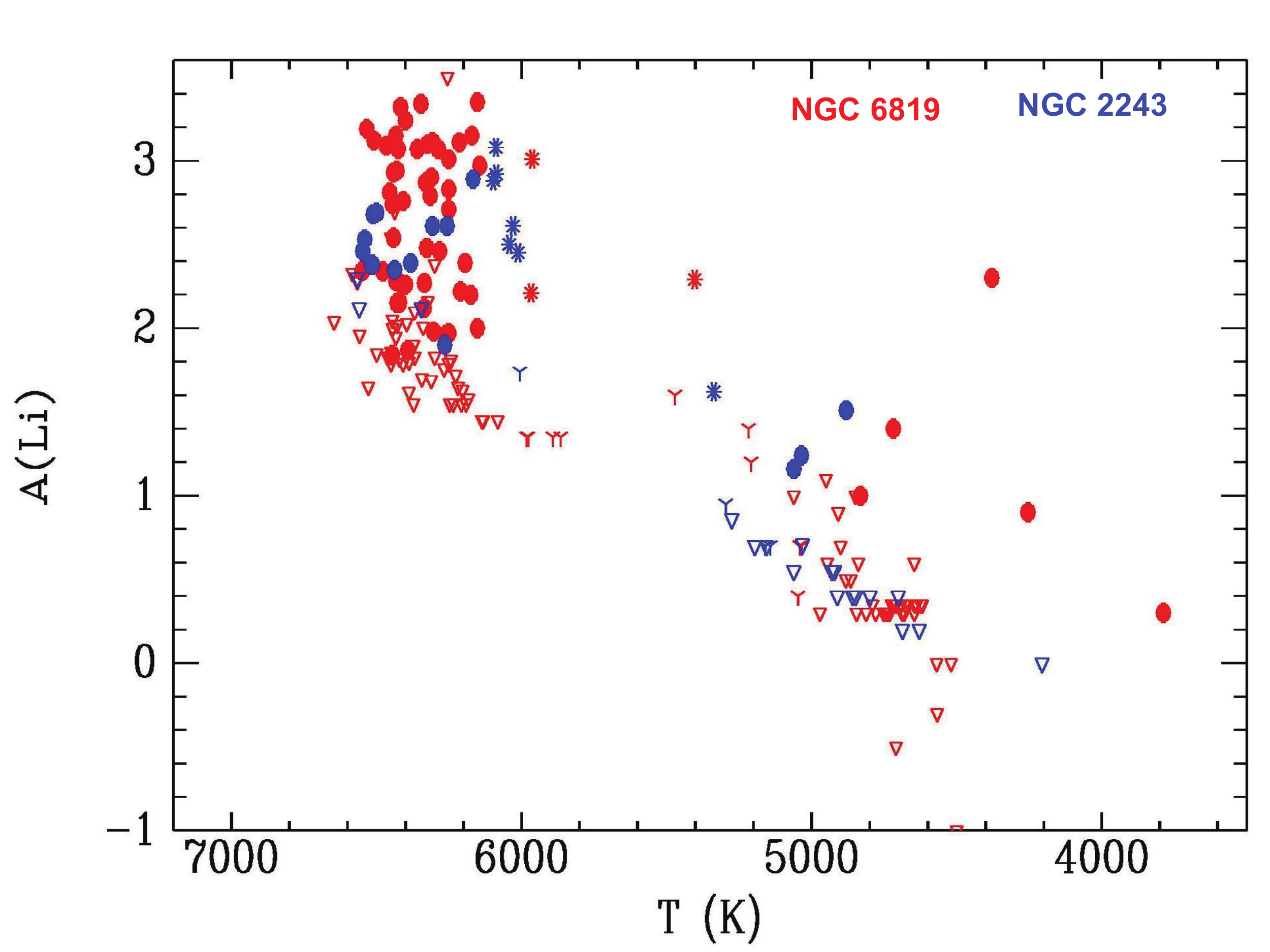}
\caption{Same as Fig. 3 for NGC2243 (blue) \& NGC6819 (red). Three-point stars are subgiants
with only upper limits. }
\end{figure} 

Symbols have the same meaning with one addition. Subgiants with only 
upper limits to A(Li) are designated by a 3-point Y. The change in 
distribution is striking: a) stars with measurable Li are now the 
exception, rather than the rule, independent of whether the star is 
a first-ascent red giant or a clump star. Even the majority of the 
subgiants have only upper limits to A(Li); b) with only one exception 
(W7017 in NGC6819), all stars again fail to meet the standard of 
Li-rich; c) while there appears to be a decline in the upper limit 
of A(Li) with decreasing \teff, this is purely due to the expected 
line strength increase as \teff\  drops, making the upper limit more 
restrictive, and has no bearing on internal evolution; d) among 
cliff dwellers, the fraction of stars with only upper limits 
to A(Li) has increased dramatically, and the majority of stars 
leaving the MS will do so with A(Li) $<$ 2.7, even before 
entering the subgiant branch. In fact, the range in A(Li) among the 
cliff stars now extends as low as the values found among stars 
within the LD, i.e. cliff dwellers are recreating the Li 
pattern long since completed by less massive stars found within 
the LD. It is probable that any additional mechanism 
that further reduces Li along the giant branch will place the 
evolved products of these future giants at a Li-level 5 to 10 times 
lower than red giants within NGC7789/2506. 

\section{Solutions \& Conclusions}
Since the red giant Li distribution for the older stars is so 
dominated by the decline in Li among stars while still at the turnoff, 
the question becomes why do these stars, which supposedly 
lack any significant degree of convection in their atmospheres, 
suddenly exhibit dramatic declines in their atmospheric Li? The 
key lies among the mechanisms often discussed when 
attempting to explain the origin of any non-convection-related Li 
decline, e.g. the origin of the LD itself. Perhaps the most 
prominent solution is tied to mixing within the atmosphere triggered 
not by convection but by stellar rotation, specifically the decline in 
rotation with time. It has been known since 
the discovery of the LD that there is a well 
delineated correlation between a star's rotation rate and the 
degree of Li-depletion, with more rapid rotators showing greater 
depletion. While this may seem like a contradiction with the 
previous claim, remember that the measured rotational 
speeds are a snapshot of the current state of the stars' atmosphere 
rather than an indicator of the original V$_{rot}$. If it 
is the spindown of  a star that triggers mixing and a reduction 
in Li, stars that have spun down from a high initial V$_{rot}$ 
to a modest value will exhibit significantly more depletion 
than a star which was rotating only moderately to begin with 
and has a minor reduction in V$_{rot}$ over time.

\begin{figure}
\figurenum{5}
\includegraphics[angle=270,width=\columnwidth]{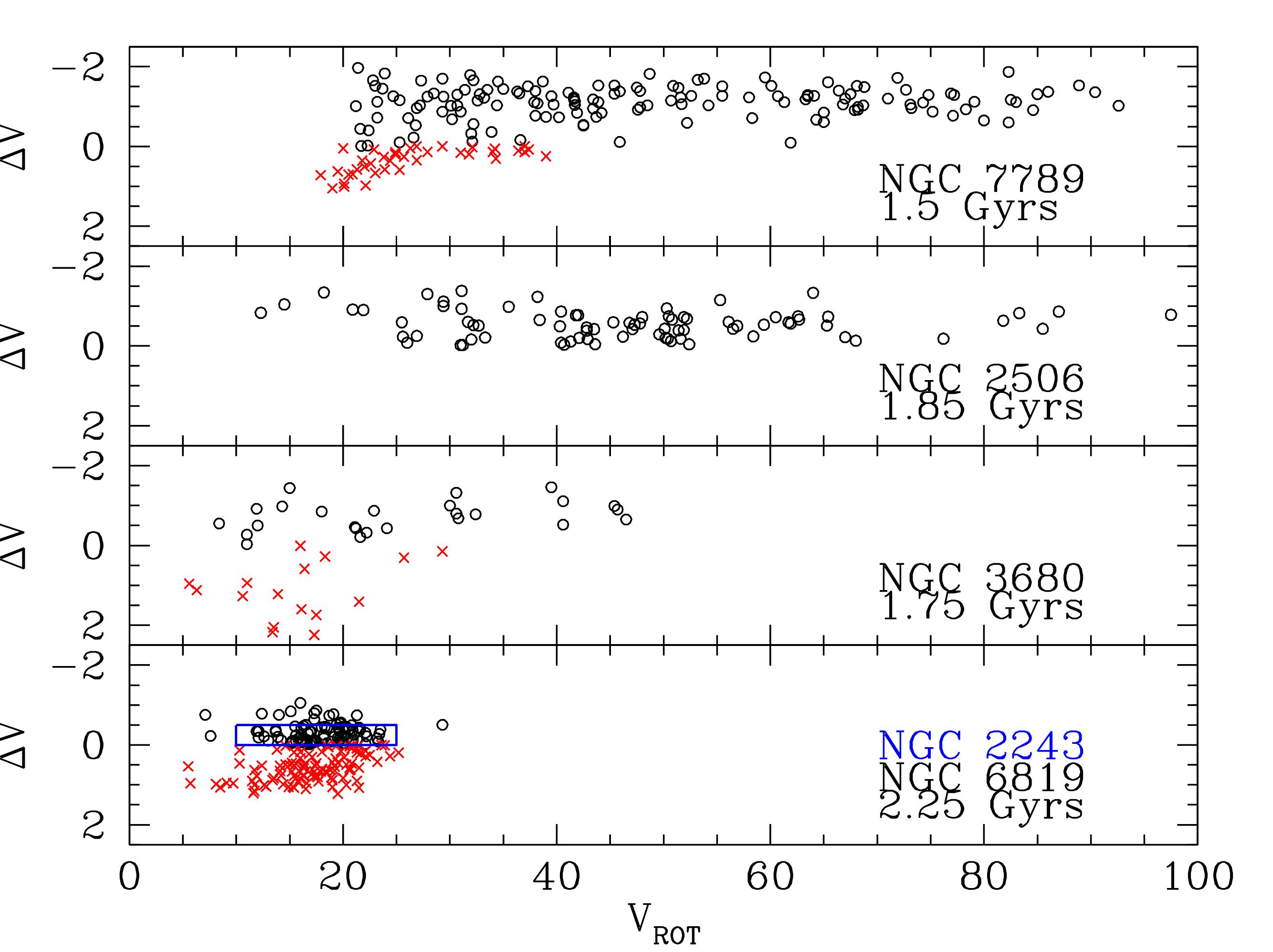}
\caption{V$_{rot}$ distribution for cliff dwellers (black circles) and LD
stars (red crosses) in each cluster. Blue box outlines the cliff distribution
for NGC2243. }
\end{figure}

To test this, we plot in Fig. 5 the V$_{rot}$ distributions for the OC, ordered by their Li state of evolution, 
with the least evolved at the top. For each OC, the dark circles 
identify only stars on the cliff, sorted by distance above the cliff, 
while the red symbols show the stars within the LD. The blue rectangle 
shows the V$_{rot}$ range for NGC2243 stars observed above the LD. 
The range is virtually identical to that found for its 
Li-phase companion, NGC6819. If one assumes that the V$_{rot}$ 
distribution in NGC6819/2243 at a younger age was comparable to 
that in NGC7789, already reduced to a smaller spread by an age 
of 1.5 Gyr, the typical star in the older OC has spun down 
by a factor of 2-4 within the last Gyr. When combined with the 
longer timescale for the H-exhaustion phase among these lower mass 
stars, we conclude that the primary source of the expanded range 
in A(Li) among the stars leaving the turnoff is mixing triggered 
by the dramatic reduction in V$_{rot}$ over the timescale of $\sim$0.5 Gyr. 
This is the same physical process that led to the creation of the 
LD among the OC when younger, i.e. during the first 0.5 
Gyr of evolution upon arrival on the zero-age MS,  
stars within the LD experienced spindown of V$_{rot}$ from their 
original state and this generated enough mixing to 
deplete Li to levels too low to measure by Hyades age.

We conclude that:
a) The high mass boundary of the LD is sharp and well-defined; 
once formed, it remains unchanged as the OC ages; b) 
the primary parameter defining the mass of the boundary 
is the metallicity; the stellar mass at the 
edge of the Li cliff decreases by $\sim$0.4 M$_{\sun}$ for each dex drop 
in [Fe/H] between [Fe/H]=+0.4 and -0.5; c) due to the absence of 
any underlying mixing mechanism, stars occupying the cliff 
initially retain the primordial cluster value, 
declining to A(Li) $<$ 1.5 only upon evolution across 
the subgiant branch to the base of the first ascent giant branch. 
There is weak evidence for a slight decline in A(Li) as the stars 
ascend the giant branch until the apparent level of the bump is 
attained and A(Li) drops below 1.0. With a dramatic decline in Li 
at He-ignition, A(Li) for He-burning stars drops so low that 
only upper limits are attainable for clump stars; d) as the mass 
range of the cliff stars declines and the LD 
is approached, an increasing fraction of turnoff stars exhibit 
severe declines in A(Li) while still at the turnoff, so extreme 
that the limiting range of A(Li) approaches that found 
within the LD. Thus, a growing fraction of the 
stars entering the more populated subgiant branch have A(Li) well 
below the canonical value (1.5) of their higher mass counterparts. 
By the time these stars reach the first ascent giant branch, 
their Li abundance is so low that only upper limits are observable 
and any additional depletions caused by the bump or He-ignition 
drive A(Li) well below that present among the higher 
mass stars; e) the underlying source of this changing pattern with declining mass 
for stars outside the LD appears to be the spindown of stars 
at the turnoff coupled with the longer time spent in this evolutionary 
phase compared to stars of higher mass. Over a timescale of $\sim$0.5 Gyr, 
the turnoff stars on the cliff spindown by a factor of 2 to 5, 
triggering atmospheric mixing that reduces Li by a factor of 4 or more; 
f) Li depletion experienced by the cliff dwellers reproduces 
the LD among stars of higher mass than the cliff boundary. This 
implies that designation as Li-rich, traditionally applied to 
any giant with A(Li) $>$ 1.5, no longer holds for these stars or for 
stars evolving from within the LD since the starting point for Li 
enhancement can be 1 to 3 orders of magnitude lower. If enrichment 
among the giants is defined by addition of a fixed amount of Li, this 
makes no difference since the amount of enriched material defines the 
observed A(Li) in the final state. If, however, the degree of 
enrichment scales with the initial A(Li), evolved giants with A(Li) 
well below 1.5 would now become classified as Li-rich, a boundary 
that will vary with both age and [Fe/H]; g) Li-rich giants aside, 
the distribution of Li among a random sample 
of giants should show a significant depletion for stars within a mass 
range that extends to a mass higher than predicted by the cliff 
mass boundary, with a specific range that varies with metallicity; 
more metal-deficient giants should have excessive Li depletion 
for lower mass giants at a greater age than metal-rich giants. 
\acknowledgements
The authors gratefully acknowledge the many colleagues who have 
contributed to this project over the years: Dianne Harmer, Kevin Croxall, 
Jeff Cummings, Donald Lee-Brown, Ryan Maderak, Qinghui Sun, Samantha 
Brunker, Evan Rich, Brian Schafer, and David Thomas.

\end{document}